\documentclass[conference]{IEEEtran}
\IEEEoverridecommandlockouts
\usepackage{cite}
\usepackage{amsmath,amssymb,amsfonts}
\usepackage{algorithmic}
\usepackage{graphicx}
\usepackage{textcomp}
\usepackage{xcolor}
\usepackage{fancyhdr}

\definecolor{arxiv}{RGB}{127,127,127}

\makeatletter
\def\ps@headings{% default to standard twoside headers, no footers
\let\@oddhead\@empty
\let\@evenhead\@empty
\def\@oddfoot{\footnotesize \textcolor{arxiv}{European Space Agency • Security for Space Systems (3S) 2024}\hfill\textcolor{arxiv}{\thepage}}%
\def\@evenfoot{\footnotesize \textcolor{arxiv}{European Space Agency • Security for Space Systems (3S) 2024}\hfill\textcolor{arxiv}{\thepage}}
}
\def\ps@IEEEtitlepagestyle{% default title page headers, no footers
\let\@oddhead\@empty
\let\@evenhead\@empty
\def\@oddfoot{\footnotesize \textcolor{arxiv}{European Space Agency • Security for Space Systems (3S) 2024}\hfill\textcolor{arxiv}{\thepage}}%
\def\@evenfoot{\footnotesize \textcolor{arxiv}{European Space Agency • Security for Space Systems (3S) 2024}\hfill\textcolor{arxiv}{\thepage}}
}
\makeatother

\def\BibTeX{{\rm B\kern-.05em{\sc i\kern-.025em b}\kern-.08em
    T\kern-.1667em\lower.7ex\hbox{E}\kern-.125emX}}

\begin{document}

\title{Security Challenges of Complex Space Applications: An Empirical Study}

\author{\IEEEauthorblockN{Tomas Paulik}
\IEEEauthorblockA{\textit{Department of Computer Science and Technology} \\
\textit{University of Cambridge}\\
United Kingdom \\
tp530@cam.ac.uk}
}

\maketitle

\begin{abstract}
Software applications in the space and defense industries have their unique characteristics: They are complex in structure, mission-critical, and often targets of state-of-the-art cyber attacks sponsored by adversary nation states. These applications have typically a high number of stakeholders in their software component supply chain, data supply chain, and user base. The aforementioned factors make such software applications potentially vulnerable to bad actors, as the widely adopted DevOps tools and practices were not designed for high-complexity and high-risk environments.\\
In this study, I investigate the security challenges of the development and management of complex space applications, which differentiate the process from the commonly used practices. My findings are based on interviews with five domain experts from the industry and are further supported by a comprehensive review of relevant publications.\\
To illustrate the dynamics of the problem, I present and discuss an actual software supply chain structure used by Thales Alenia Space, which is one of the largest suppliers of the European Space Agency. Subsequently, I discuss the four most critical security challenges identified by the interviewed experts: Verification of software artifacts, verification of the deployed application, single point of security failure, and data tampering by trusted stakeholders. Furthermore, I present best practices which could be used to overcome each of the given challenges, and whether the interviewed experts think their organization has access to the right tools to address them. Finally, I propose future research of new DevSecOps strategies, practices, and tools which would enable better methods of software integrity verification in the space and defense industries.
\end{abstract}

\begin{IEEEkeywords}
Cybersecurity, software development life cycle, DevOps, DevSecOps, code injection, distributed systems, microservices, blockchain.
\end{IEEEkeywords}

\section{Introduction} \label{intro}
Software development life cycle (SDLC) is a process which covers all the stages of software: formation, requirements planning, design and construct, testing, and product release \cite{r1}\cite{r2}. The last three of these stages can be integrated and automated using specialized tools and practices for \textit{development} and IT \textit{operations}, otherwise known as DevOps. These tools and practices can be viewed as a progression of Agile methodologies, which emphasize working software, collaboration, speed, and fast response to change \cite{r3}. Formally, we can define DevOps as a “collaborative and multidisciplinary effort within an organization to automate continuous delivery of new software versions, while guaranteeing their correctness and reliability” \cite{r4}.

DevOps are currently a widely adopted practice in the software industry \cite{r5}\cite{r6}. The adoption is primarily driven by the gains in the business value DevOps can deliver \cite{r7}. However, the automated continuous delivery of software has created new challenges for the DevOps practitioners. One such challenge is keeping the security of the resulting software artifacts while maintaining the agile nature of the software development process \cite{r8}.

The concept of putting emphasis on security within the DevOps process has led to the creation of a spin-off term, DevSecOps, meaning \textit{development}, \textit{security}, and \textit{operations}. In academic literature, there is no single definition of DevSecOps, although generally, the term refers to security practices and testing being performed early in the SDLC \cite{r9}.

OWASP DevSecOps Maturity Model is a popular open-source theoretical framework which defines security measures that could be applied when using DevOps strategies, and the ways these measures can be prioritized \cite{r10}. It recognizes five levels of security measures (i.e., levels 1 to 5), available along five security dimensions: Build and Deployment, Culture and Organization, Implementation, Information Gathering, and Test and Verification. Each of the dimensions consists of multiple security practices, where each practice is associated with one security level.

Even though OWASP DevSecOps Maturity Model and similar frameworks are comprehensive, as I present later in the study, there is a need for further research of DevSecOps strategies, practices and tools which would enable stricter methods of software integrity verification.

To introduce the context of the DevSecOps practices in the space and defense industries, I present below an actual software supply chain structure used by Thales Alenia Space (TAS). The company is a provider of space-based systems, including satellites and ground segments, used for navigation, telecommunications, earth observation, space exploration, and scientific research. TAS is also one of the key technical contributors to Galileo—a global satellite navigation system operated by the European Union Agency for the Space Programme \cite{r_tas}.

\begin{figure*}[htbp]
    \centering
    \includegraphics[width=1\linewidth]{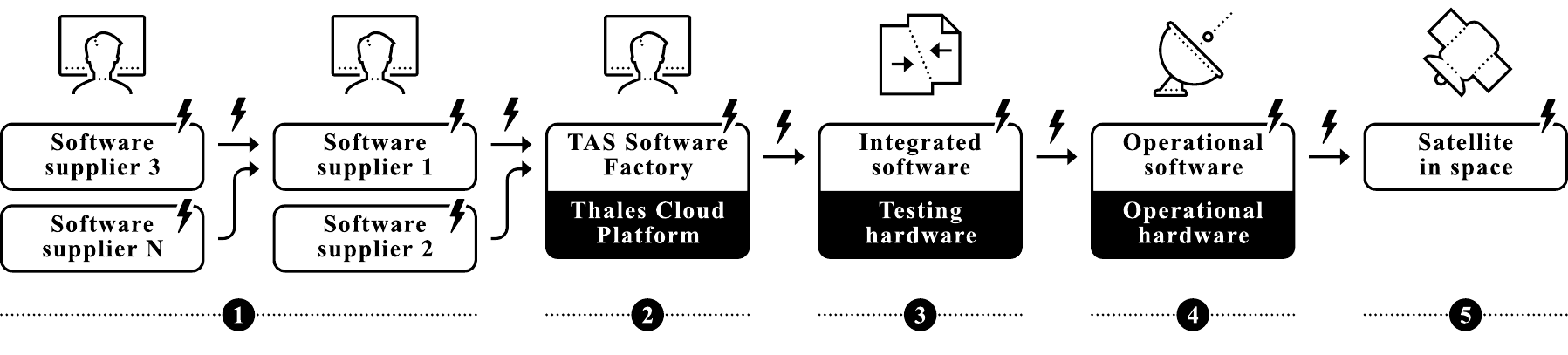}
    \caption{An actual software supply chain structure used by Thales Alenia Space for ESA programs such as Galileo, Copernicus, and IRIS\textsuperscript{2}.}
    \label{fig}
\end{figure*}

Fig. 1, published with permission of TAS, illustrates the entire flow of software from third-party component suppliers to the mission center and satellites in space. The following are the descriptions of the corresponding segments of Fig. 1.

\begin{enumerate}
\item \textbf{Software suppliers}: Third-party companies developing building blocks such as components, libraries, or entire microservices \cite{r25}. The suppliers usually form a multi-level structure (i.e., supply chain).
\item \textbf{TAS Software Factory}: The company’s internal development department responsible for the assembly of the final software deliverables for the end-client, such as the European Space Agency (ESA). TAS uses their own cloud infrastructure for the DevSecOps purposes.
\item \textbf{Integrated software}: After the final assembly, the software is subject to rigorous in-house integration testing (at TAS), with the utilization of a dedicated testing hardware.
\item \textbf{Operational software}: The software deployed on the operational infrastructure of the customer. Depending on the project, such infrastructure might include ground stations, satellite control center, mission center, launch site (i.e., spaceport), and satellite manufacturing facility.
\item \textbf{Satellite in space}: A device equipped with its own application software and controlled by Operational software.
\end{enumerate}

Lightning icons in Fig. 1 represent points of possible security vulnerabilities, as identified by TAS. Attack vectors which could be applied to these points include malicious code injection, insider attack, and man-in-the-middle attack.

Below are the main points of difference between common commercial software supply chains and the supply chain outlined in Fig. 1:

\begin{enumerate}
\item ESA space programs developed by TAS and their suppliers include Galileo (satellite navigation), Copernicus (earth observation), and IRIS\textsuperscript{2} (Infrastructure for Resilience, Interconnectivity and Security by Satellite). For adversary nation states, these programs present high-value targets and are subject to state-of-the-art cyber attacks.
\item Programs listed in the previous point are subject to legal requirements, as some aspects of these programs might have military applications and could be a matter of national security. Additional requirements might be imposed by the funding body or the development consortium itself. Furthermore, the participating companies might have their own internal requirements, as some of them might be regulated by their respective jurisdictions.
\item There is no universally adopted public key infrastructure (PKI) among the European software suppliers in the space industry, which would enable them to use digital certificates and public key cryptography to sign their software artifacts \cite{r_pki}. Such PKI could be used for the validation of integrity and authenticity of software packages. Even though a universally adopted PKI would be an improvement to the current situation, sections \ref{sc1} and \ref{sc2} of this study discuss practical limitations of such PKI.
\item Due to the previous two points and due to confidentiality reasons, the companies participating in the supply chain typically cannot use managed package repositories such as npm or MyGet.
\item Segments 4 and 5 of Fig. 1 usually consist of a high number of software applications of different verticals, ranging from containerized microservices (written in high-level programming languages) to embedded applications (written in low-level programming languages). As a result, software inspection and integration (which is illustrated in segment 2 of Fig. 1) is harder compared to traditional commercial software projects, and requires a team of computer scientists of different backgrounds.
\item Deep supply chains where some of the software artifacts are in a binary format are highly vulnerable to bad actors, especially if the malicious code is injected at the early levels of the supply chain. The deeper the malicious code is encapsulated under the layers of deliverables, the harder it is to discover such code.
\end{enumerate}

As for the research gap, limited research has been conducted on the specific needs of the development teams working in space and defense industries, and whether the conventional SDLC process and DevSecOps practices cover the security needs of such teams.

The goal of this study is to investigate the real-world security challenges of the SDLC of complex space applications, which differentiate the process from the traditional SDLC. Given the high sensitivity of the matter, I have conducted anonymized interviews with five problem domain experts. The reason for anonymity was to enable the experts to speak freely about the security challenges they are aware of at their organization, or at their supplier and partner organizations. During the interviews, the experts were not presented with a predefined list of security challenges. Instead, they were encouraged to present their own views. After the interviews, I analyzed and aggregated the answers, and produced a set of four top-most challenges, which are presented in this study. Each of the challenges is accompanied by a discussion supported by conventional literature review.

This study aims to address the following research questions:
\begin{itemize}
\item \textbf{RQ1}: What security challenges of the SDLC process of complex space applications pose the highest risk for the organizations using such applications?
\item \textbf{RQ2}: For each of the security challenges, do the organizations developing such applications have the right tools to address the given challenge?
\end{itemize}

The motivation of conducting this study is to openly present the academics, researchers, and professionals in the space and defense community a holistic view of the real-world security challenges of their SDLC processes, in contrast to keeping the discussion centered around theoretical frameworks such as the OWASP DevSecOps Maturity Model. Specifically, the goal of this research is to help the space and defense communities in the following areas:
\begin{enumerate}
\item To understand the main SDLC security challenges present in their industry.
\item To increase the level of awareness (at the given organizations) of the importance of performing a secure SDLC process.
\end{enumerate}

The rest of the document is organized as follows: Section \ref{methodology} outlines the research methodology. Section \ref{profiles} presents the profiles of the interview participants. Section \ref{validity} lists the possible threats to validity. Section \ref{results} presents the research results and discussion. Section \ref{summary} provides a summary of the interviews. Finally, the conclusion and future work are presented in section \ref{conclusion}.

\section{Research Methodology} \label{methodology}

The research presented in this study was carried out using the steps below:

\begin{enumerate}
\item \textbf{Initial literature review}: A conventional literature review was performed to gain a detailed insight into current trends and topics in the field of DevSecOps.
\item \textbf{Definition of the research questions}: The knowledge generated by the initial literature review was used to design the research questions of this study.
\item \textbf{Interview design}: The interviews were intentionally structured such that they enabled the interviewees to provide long answers to both questions. The aim was to gain as many details as possible.\\
With the first research question (RQ1), the interviewees were asked to identify and rank five top-most security challenges with the risk level ranging from 1 to 5, where 5 represents the highest risk. The interviewees were allowed to use each risk level only once. They were also allowed to change their previously assigned risk levels by the end of the interview.\\
The second research question (RQ2) was asked as a simple Yes/No question, however, the interviewees were given the opportunity to further justify their answer.
\item \textbf{Sampling of the interviewees}: Five domain experts we selected for the interview, such that they represent an entire spectrum of the given industrial segment and its wider ecosystem. The interview participants are of diverse backgrounds, and work at organizations ranging from tech startups to multinational conglomerates.
\item \textbf{Analysis of the interview responses}: The transcripts of the responses were analyzed using the Content Analysis methodology \cite{r11}. This methodology was used to examine patterns in the interviews in a systematic manner.
\item \textbf{Selection of the most critical challenges}: The sums of the risk levels assigned by the interviewed experts were used to score the security challenges. The challenges with the highest overall score were selected for further investigation. However, given that the interviewees were allowed to name any security challenges (in contrast to choosing them from a predefined list), only four challenges were independently identified by at least two experts, where the overall score exceeded 5.
\item \textbf{Further literature review}: A conventional literature review was performed to gain further knowledge about the top four security challenges. The new knowledge, combined with the insights gained during the interviews, is presented in section \ref{results}.
\end{enumerate}

\section{Profiles of The Interview Participants} \label{profiles}

Table I contains an anonymized overview of the profiles of the interview participants.

\begin{table}[htbp]
\caption{Profiles of the Interview Participants}
\begin{center}
\begin{tabular}{ | p{0.3\linewidth} | p{0.4\linewidth} | p{0.12\linewidth} | }
\hline
\textbf{Role within the organization} & \textbf{Organization type} & \textbf{Region} \\
\hline
Chief technology officer & Software startup in the space industry with less than 50 employees. & EU \\
\hline
{Chief product security officer} & {Space technology manufacturer with less than 10,000 employees.} & {EU} \\
\hline
Technical officer & Intergovernmental organization devoted to space activities with less than 3,000 employees. & Europe and North America \\
\hline
Senior-level software engineer & Multinational aerospace corporation with less than 150,000 employees. & EU \\
\hline
Software architect & Developer of industrial solutions and simulation software in the field of aeronautics with less than 2,000 employees. & EU \\
\hline
\end{tabular}
\label{tab1}
\end{center}
\end{table}

\section{Threats to Validity} \label{validity}
This study is subject to the following threats to validity:
\begin{enumerate}
\item \textbf{Construct validity}: The final set of the security challenges extracted from the interviews and their order may be influenced by the personal bias of the author and the objectivity of the presented set cannot be verified. The prioritization of the challenges depends on the personal opinions of the interviewed experts. 
\item \textbf{External validity}: The presented results cannot be generalized as the study has been conducted with only five participants.
\end{enumerate}

\section{Research Results and Discussion} \label{results}

The following four security challenges have been identified by the interviewed experts as the most critical for a secure SDLC process of complex space applications (RQ1). The challenges are ordered by risk level, starting with the highest one. Each of the sections below ends with an aggregated view of the interviewed experts on whether their organization has the right tools to address the given security challenge (RQ2).

\subsection{Verification of Software Artifacts} \label{sc1}
As discussed in section \ref{intro} of this study, the software applications in the space and defense industries are built using a wide variety of programming languages and frameworks, depending on their vertical and use. While applications running on satellites are typically written in languages such as C, C++, and Rust, control center and mission center applications are predominantly written in dynamic programming languages. Both of these groups of languages can be subject to a malicious code injection performed by a sophisticated attacker, however, it is the latter that is particularly vulnerable to such attack.\\
Dynamic programming languages such as JavaScript and Python are consistently among the most popular in the world by almost any measure \cite{r12}. Frameworks based on these languages, including React \cite{r13}, Node.js \cite{r14}, and PyTorch \cite{r15} are leaders in their respective application verticals: Front-end applications, API services, and machine learning applications. Ironically, both programming languages were designed in early 1990s as easy-to-learn languages aimed towards hobbyists and semi-professional developers, prioritizing simplicity of use over professional features and practices \cite{r16}\cite{r17}. The heritage of such philosophy can be seen even three decades later—applications (and software packages) produced in both languages are usually deployed to production (or to the package repository) without compilation into a machine code (or other binary format, such as Bytecode or MSIL). To illustrate the nature of the problem, let’s consider the following scenario:

\begin{enumerate}
\item A developer at a fictional company Supplier 1 is working on a software application (Application 1) written in JavaScript or TypeScript programming language.
\item The developer clones a git repository with Application 1’s source code to their computer.
\item The git system of Supplier 1 uses two-factor authentication, branch permissions, and other best practices in terms of software security.
\item The repository is cloned on the developer’s computer. The developer runs the “npm install” command to restore the project’s dependencies.
\item A sophisticated attacker who has access to the developer’s file system modifies one of the restored JS files and adds malicious code. (Alternatively, the attacker could use a man-in-the-middle network attack to change the content of one of the dependencies during the restoration.)
\item The developer writes some new code and creates a pull request. The malicious code added by the attacker stays invisible, as the “node\_modules” folder containing the dependencies is listed in the “.gitignore” file.
\item A co-worker of the developer reviews the pull request and approves it (which is perfectly reasonable, as the malicious code is not included in the commit due to the presence of the “.gitignore” file). All the unit tests and static code analysis tests pass. The new code is now part of the main development branch. Git commits internally use SHA-1 hashes to ensure the consistency of the content.
\item The developer switches their local git repository to the main branch as they intend to build the application. For additional security, the developer executes “npm audit signatures”. The audit completes successfully, all the packages have verified registry signatures.
\item The developer builds the application using “npm run build”. At this point, the malicious code is included in the build output. The “package-lock.json” file contains SHA-512 hash codes of the modules (one of which is altered), although the integrity of the modules is not verified at this stage. (Alternatively, the steps 5 to 9 can be applied to a different pipeline where the build artifacts are produced by a CI/CD build agent. In such a case, the attacker needs to get access to the computer which hosts the build agent. Building and testing of artifacts in virtual environments is recommended by the OWASP DevSecOps Maturity Model.)
\item The developer uploads the assembled version of Application 1 to a secure cloud storage. With the belief of enhancing the security further, the developer manually creates a hash code of Application 1 and sends it to the customer organization (Supplier 2) up in the supply chain.
\item A developer working at Supplier 2 downloads Application 1 from the secure storage. The download is seamless, performed via HTTPS connection with the use of a strong cryptographic protocol TLS version 1.3. The developer at Supplier 2 manually verifies the hash code provided by Supplier 1. The hashes are identical, everything looks correct.
\item Supplier 2 creates a new version of their own application, Application 2, and creates an application bundle consisting of both applications. The application bundle is sent up in the supply chain. At this stage, the infected code is buried too deeply to be discovered.
\item A perfect illusion of security is preserved throughout the entire process. DevSecOps practices are in place at all the participating organizations. And yet, their code base is infected.
\end{enumerate}

The OWASP DevSecOps Maturity Model specifically recommends “Signing of artifacts” in their “Build and Deployment” dimension of level 5. However, when exploring the recommendation further, it only refers to git commit signing by the author. The recommendation is not covering signing (and integrity verification) of dependencies and build outputs. Generally, it would be difficult to make such a recommendation in a theoretical framework as, at the time of the writing, there is no publicly available tool or service which would provide functionality comprehensive enough to prevent the code injection attack vector outlined above.

All the three interviewed experts who have identified the challenge think their organization doesn’t have the right tools to address the challenge. One interviewee has stated that this challenge presents for their organization a major issue, because the inability to trace the origin of a compromised software artifact means that the supplier responsible for the breach of security cannot be held accountable.

\subsection{Verification of The Deployed Application} \label{sc2}

A successful deployment of an application to a remote server might be the last step of the DevOps process, although it is not the end of the application’s life cycle \cite{r2}. Let’s consider a scenario where we have an API application written in Python and FastAPI \cite{r18}, deployed on a virtual machine (VM), and hosted in a cloud environment. Even if the application is meant to continuously run on the server from the deployment until the release of its next version, chances are that the application will be occasionally terminated and started again. This is because the VM might be restarted due to the technical needs of the underlying infrastructure, or due to an operating system update. A technically advanced attacker might utilize the opportunity to modify the code of the application during the start of the operating system. Such change would be almost unnoticeable using the common application health monitoring practices.

A seemingly straightforward way of solving the problem would be to use Code Signing – a technique to confirm the software author and guarantee that the code has not been altered or corrupted since it was signed \cite{r19}. However, the interviewed experts have confirmed that there is no public key infrastructure (or an equivalent mechanism) across the supply chain of European software companies, which would enable them to practice Code Signing. According to the interviewed experts, the absence of the public key infrastructure is largely a non-technical problem, of solving which would require a lot of business and political negotiations. Regardless of the non-technical aspects, Code Signing as a technique could be insufficient for mission-critical applications. Academic literature provides several articles on “certified malware” \cite{r20}, illegal trading of signing certificates \cite{r21}, and abuses of the authentication mechanism \cite{r22}.

Much like the challenge outlined in section \ref{sc1}, the verification of deployed applications on the level of the operating system is, at the time of the writing, a difficult open problem with no clear best practices or technical consensus among the European software suppliers in the space and defense industries.

Two out of three interviewed experts who have identified the challenge think their organization doesn’t have the right tools to address the challenge.

\subsection{Single Point of Security Failure}

The idea behind this challenge is that a single vulnerability of a small software component can compromise the security of a large software infrastructure. This phenomenon is well-studied, and the literature provides a variety of mitigation strategies to address the risk \cite{r23}\cite{r24}. Even though the idea is not new, it is understandable that the interviewed experts are concerned about the potential business damages caused by this scenario.

A general best practice to address the risk is to design the application as a set of microservices during the design stage of the SDLC, in contrast to choosing the monolithic architecture \cite{r25}. Microservices are an architectural pattern which divides the application into multiple loosely coupled, fine-grained services, communicating through a lightweight protocol. The advantage of the architecture is that each of the smaller services is easier to test, maintain, and scale \cite{r26}. From the security perspective, compromising the security of a single microservice doesn’t automatically affect the other microservices of the application, although depending on the vulnerability, the problematic code component might be present in other microservices.

The OWASP DevSecOps Maturity Model recommends using the “Microservice-architecture” in their “Implementation” dimension of level 5, stating that “monolithic applications are hard to test.”

All the three interviewed experts who have identified the challenge think their organization does have the right tools to address the challenge. One of the experts has expressed their view that despite the availability of the right tools, the challenge is one of the highest their organization is facing.

\subsection{Data Tampering by Trusted Stakeholders}

Companies in the space and defense industries are typically large (in terms of the number of employees), have deep supply chains, and deal with expensive products and components. Manufacturing defects, failed component tests, and errors in technical documents can translate into business damages worth millions of Euros. Therefore, when such damage happens, the responsible employees have a very strong incentive to obfuscate their responsibility and move it to a different department within the company, or up or down the supply chain. Such obfuscation can be achieved, for example, by manually changing the records on the company’s database server (by bypassing the application logic and directly modifying the data). These practices are criminal in nature, and lead to disputes, manufacturing delays, and ultimately to financial losses for the business.

From the technical perspective, this problem can be addressed by a use of a different type of the underlying database system \cite{r27}. Instead of using a traditional database such as relational database (SQL) or non-relational database (NoSQL), businesses could use a distributed database system based on a shared immutable ledger. Such a category of systems is commonly referred to as blockchain databases \cite{r28}.

When designing a blockchain-based application, businesses can choose either to use one of the publicly available blockchain infrastructures, or create their own network of nodes, where each of the nodes could be, for example, operated by one of the organizations in the given supply chain. As a result, businesses could utilize blockchain technology to increase transparency, trust, and traceability in their own organization, or within their supply chain.

The two interviewed experts who have identified the challenge both think their organization does have the right tools to address the challenge. One of the experts has mentioned that even though the right tools are available, their organization is slow in adopting them.

\section{Summary of the Interviews} \label{summary}

Table II shows a summary of the identified security challenges, including their risk levels assigned by the interviewed experts (RQ1). Table III provides an overview of whether the interviewed experts think their organization has the right tools to address the identified security challenges (RQ2).

\begin{table}[htbp]
\caption{Summary of the Identified Security Challenges}
\begin{center}
\begin{tabular}{ | p{0.3\linewidth} | p{0.4\linewidth} | p{0.12\linewidth} | }
\hline
\textbf{Security challenge} & \textbf{Risk levels assigned by the experts} & \textbf{Overall score} \\
\hline
Verification of software artifacts & 5, 5, 4 & 14 \\
\hline
Verification of the deployed application & 5, 4, 4 & 13 \\
\hline
Single point of security failure & 4, 3, 2 & 9 \\
\hline
Data tampering by trusted stakeholders & 5, 3 & 8 \\
\hline
\end{tabular}
\label{tab1}
\end{center}
\end{table}

\begin{table}[htbp]
\caption{Overview of whether the interviewed experts think their organization has the right tools to address the identified security challenges}
\begin{center}
\begin{tabular}{ | p{0.3\linewidth} | p{0.26\linewidth} | p{0.26\linewidth} | }
\hline
\textbf{Security challenge} & \textbf{Organizations with the right tools} & \textbf{Organizations without the right tools} \\
\hline
Verification of software artifacts & 0 & 3 \\
\hline
Verification of the deployed application & 1 & 2 \\
\hline
Single point of security failure & 3 & 0 \\
\hline
Data tampering by trusted stakeholders & 2 & 0 \\
\hline
\end{tabular}
\label{tab1}
\end{center}
\end{table}

\section{Conclusion and future work} \label{conclusion}

In this study, I presented the four most critical security challenges identified by the interviewed experts: Verification of software artifacts, verification of the deployed application, single point of security failure, and data tampering by trusted stakeholders. These security challenges can be divided into two categories:

\begin{enumerate}
\item The first two challenges are both difficult open problems, and the interviewed experts stated their organizations don’t have the right tools to address them.
\item The second two challenges are well-studied problems; the interviewed experts think their organizations have the right tools to address them, although dealing with these challenges is hard from both technical and organizational perspectives.
\end{enumerate}

As a part of my future work, I would like to further focus on the findings related to the first two challenges.

As discussed in sections \ref{sc1} and \ref{sc2}, complex supply chains of software components in the space and defense industries create a potential opportunity for bad actors to perform advanced attack vectors involving alteration or replacement of software artifacts. A pre-requirement of such an attack vector is the attacker’s ability to access the file system on the developer’s computer, which, on its own, already means the security of the targeted organization has been compromised.

I would like to conduct further research on DevSecOps strategies, practices and tools which assume in advance that one or more developer computers could be compromised. Such a paradigm would require advanced build tools, much stricter methods of software integrity verification, and parallel assembly of software artifacts.

\section*{Acknowledgment}

I would like to acknowledge the generous support provided by TAS and all the interviewed experts, who were keen to openly talk about cybersecurity practices at their organizations.


\begin{thebibliography}{00}

\bibitem{r1} N. B. Ruparelia, “Software development lifecycle models,” ACM SIGSOFT Software Engineering Notes, vol. 35, issue 3, pp. 8--13, May 2010.

\bibitem{r2} M. Paul, “Official (ISC)\textsuperscript{2} Guide to CSSLP CBK,” Boca Raton, Fl: CRC Press/Taylor \& Francis Group, 2014.

\bibitem{r3}H. B. Christensen, “Teaching DevOps and Cloud Computing using a Cognitive Apprenticeship and Story-Telling Approach,” Proceedings of the 2016 ACM Conference on Innovation and Technology in Computer Science Education, July 2016.
‌
\bibitem{r4}L. Leite, C. Rocha, F. Kon, D. Milojicic, and P. Meirelles, “A Survey of DevOps Concepts and Challenges,” ACM Computing Surveys, vol. 52, no. 6, pp. 1--35, Jan. 2020.
‌
\bibitem{r5}M. Krey, A. Kabbout, L. Osmani, and A. Saliji, “DevOps Adoption: Challenges \& Barriers,” Proceedings of the 55\textsuperscript{th} Hawaii International Conference on System Sciences, January 2022.

\bibitem{r6}N. Forsgren, D. Smith, J. Humble, and J. Frazelle, “2019 Accelerate State of DevOps Report,” DORA \& Google Cloud, 2019.

\bibitem{r7} L. Riungu-Kalliosaari, S. Mäkinen, L. E. Lwakatare, J. Tiihonen, and T. Männistö, “DevOps Adoption Benefits and Challenges in Practice: A Case Study”, Proceedings of the 2016 Springer Conference on Product-Focused Software Process Improvement, November 2016.

\bibitem{r8} H. Myrbakken, and R. Colomo-Palacios, “DevSecOps: A Multivocal Literature Review,” Proceedings of the 2017 Springer Conference on Software Process Improvement and Capability Determination, October 2017.

\bibitem{r9} V. V. Sehgal, “Implementing DevSecOps Practices,” Packt Publishing, December 2023.
‌
\bibitem{r10} R. Brasoveanu, Y. Karabulut, and I. Pashchenko, “Security Maturity Self-Assessment Framework for Software Development Lifecycle,” Proceedings of the 17\textsuperscript{th} International Conference on Availability, Reliability and Security (ARES 2022), August 2022.

\bibitem{r_tas} J-L. Issler, A. de Latour, L. Ries, L. Lestarquit, and M. Grondin, J. Dantepal, “Lessons Learned from the use of GPS in Space: Application to the Orbital use of GALILEO,” Proceedings of the 21\textsuperscript{st} International Technical Meeting of the Satellite Division of The Institute of Navigation (ION GNSS 2008), pp. 719--735, September 2008.

\bibitem{r25} Y. Romani, O. Tibermacine, and C. Tibermacine, “Towards Migrating Legacy Software Systems to Microservice-based Architectures: a Data-Centric Process for Microservice Identification,” Proceedings of the 2022 IEEE 19\textsuperscript{th} International Conference on Software Architecture Companion (ICSA-C), pp. 15--19, 2022.

\bibitem{r_pki} ors:
J. A. Buchmann, E. Karatsiolis, and A. Wiesmaier, “Introduction to Public Key Infrastructures,” Springer, 2013.

\bibitem{r11}K. Krippendorff, “Content Analysis: An Introduction to Its Methodology,” 4\textsuperscript{th} edition, SAGE Publications, 2018.
‌
\bibitem{r12} D. Lu, J. Wu, Y. Sheng, P. Liu, and M. Yang, “Analysis of the popularity of programming languages in open source software communities,” Proceedings of the 2020 International Conference on Big Data and Social Sciences (ICBDSS), August 2020.
‌
\bibitem{r13} A. Boduch, R. Derks, and M. Sakhniuk, “React and React Native,” 4\textsuperscript{th} edition, Packt Publishing, 2022.
‌
\bibitem{r14} B. A. Syed, “Beginning Node.js,” Apress, December 2014.

\bibitem{r15} E. Stevens, L. Antiga, and T. Viehmann, “Deep learning with PyTorch,” Manning Publications, 2020.
‌
\bibitem{r16} M. Selakovic, and M. Pradel, “Performance Issues and Optimizations in JavaScript: An Empirical Study”, Proceedings of the 38\textsuperscript{th} International Conference on Software Engineering (ICSE), pp. 61--72, 2016.

\bibitem{r17} L. Jun, and L. Ling, “Comparative research on Python speed optimization strategies,” Proceedings of the 2010 International Conference on Intelligent Computing and Integrated Systems, pp. 57--59, 2010.

\bibitem{r18} B. Lubanovic, “FastAPI: Modern Python Web Development,” O’Reilly Media, November 2023.

\bibitem{r19} D. Cooper, A. Regenscheid, M. Souppaya, C. Bean, M. Boyle, D. Cooley, M. Jenkins, “Security Considerations for Code Signing,” NIST Cybersecurity White Paper, January 2018.

\bibitem{r20} D. Kim, B. J. Kwon, and T. Dumitraş, “Certified Malware: Measuring Breaches of Trust in the Windows Code-Signing PKI,” Proceedings of the 2017 ACM SIGSAC Conference on Computer and Communications Security (CCS), pp. 1435--1448, 2017.

\bibitem{r21} K. Kozák, B. J. Kwon, D. Kim, and T. Dumitraş, “Issued for Abuse: Measuring the Underground Trade in Code Signing Certificates”, arXiv:1803.02931v3 [cs.CR], February 2019.

\bibitem{r22} P. Kotzias, S. Matic, R. Rivera, and J. Caballero, “Certified PUP: Abuse in Authenticode Code Signing,” Proceedings of the 22\textsuperscript{nd} ACM SIGSAC Conference on Computer and Communications Security (CCS), pp. 465--478, 2015.

\bibitem{r23} J. Ransome, A. Misra, and B. Schoenfield, “Core Software Security: Security at the Source,” Boca Raton, Fl: CRC Press/Taylor \& Francis Group, 2014.
‌
\bibitem{r24} M. Dowd, J. McDonald, and J. Schuh, “The Art of Software Security Assessment: Identifying and Preventing Software Vulnerabilities,” Addison Wesley Professional, November 2006.

\bibitem{r26} R. Mitra, and I. Nadareishvili, “Microservices Up \& Running: A Step-by-Step Guide to Building a Microservices Architecture,” O’Reilly Media, December 2020.
‌
\bibitem{r27} C. Coronel, and S. Morris, “Database Systems: Design, Implementation, and Management,” 14\textsuperscript{th} edition, Cengage Learning, 2023.

\bibitem{r28} T. Ahram, A. Sargolzaei, S. Sargolzaei, J. Daniels, and B. Amaba, “Blockchain Technology Innovations,” Proceedings of the 2017 IEEE Technology \& Engineering Management Conference (TEMSCON), pp. 137--141, 2017.

\end{thebibliography}
\end{document}